\documentstyle[sprocl,axodraw,epsfig,abbreviations]{article}
\def\beq{\begin{equation}}
\def\eeq{\end{equation}}
\def\bea{\begin{eqnarray}}
\def\eea{\end{eqnarray}}
\def\barr{\begin{array}}
\def\earr{\end{array}}
\newcommand{\tenrm}{\mbox{}}

\begin{document}

\title{\boldmath 
SELECTRON SEARCHES IN \ee, \ep\ AND \pp\ SCATTERING
\unboldmath}

\author{FRANK CUYPERS}

\address{{\tt cuypers@pss058.psi.ch}\\
	Max-Planck-Institut f\"ur Physik,
	F\"ohringer Ring 6,
	D-80805 M\"unchen,
	Germany}


\maketitle\abstracts{
We review the selectron production mechanisms
in the \lc\ modes 
which do not require positrons.
The \sm\ backgrounds
can be rendered harmless or even be nearly eliminated
with polarized beams.
We insist on the complementarity of these different experiments.
}

\section{Introduction}

Supersymmetry is an excellent example 
for illustrating the complementarity 
of the different operating modes of a \lc.
We consider here in turn the production of \sel s
in \ee, \ep\ and \pp\ collisions \cite{ee,ep,pp}
within the framework of the \mssm.

Polarization of both incoming beams
is a major asset.
We assume here
that electron beams can be polarized up to 90\%\,
and we use realistic polarization and energy spectra \cite{ginzburg}
for the photon beams.
By design,
to avoid photon rescattering and pair production,
the photon energy spectrum 
is bound from above to about 80\%\ of the electron beam energy.
Similarly,
it is effectively bound from below to about 50\%,
because less energetic photon are produced at too wide angles
to interact efficiently.

The produced \sel s 
are typically long-lived
and eventually decay by weak interactions.
In the cases of \ep\ and \pp\ 
(as well as \pe)
collisions,
hefty \trm\ or energy cuts have to be imposed on the final state electrons,
in order to separate the \susic\ 
signal from the otherwise overwhelming \sm\ backgrounds.
One therefore concentrates in these cases 
on the simplest decay mode of the \sel\
into an electron and the lightest \no,
which is stable 
and escapes detection:
\begin{equation}
        \tilde e^-\quad\to\quad e^-\No~.
\label{seldec}
\end{equation}
In contrast,
in \ee\ collisions
no such cuts are needed to separate the signal from the background
and more complicated cascade decays can be observed \cite{casc}.

\section{\boldmath\ee\ Scattering \protect\cite{ee}}

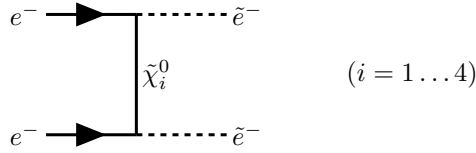
\begin{figure}[htb]
\unitlength.8mm\SetScale{2.269}
\begin{center}
\begin{picture}(50,20)(0,0)
\ArrowLine(0,0)(15,0)
\Text(-1,0)[r]{\normalsize $e^-$}
\ArrowLine(0,20)(15,20)
\Text(-1,20)[r]{\normalsize $e^-$}
\Line(15,20)(15,0)
\Text(16,10)[l]{\normalsize $\tilde\chi^0_i$}
\DashLine(15,20)(30,20){1}
\Text(31,20)[l]{\normalsize $\tilde e^-$}
\DashLine(15,0)(30,0){1}
\Text(31,0)[l]{\normalsize $\tilde e^-$}
\Text(50,10)[lc]{\normalsize $(i=1\dots4)$}
\end{picture}
\end{center}
\caption{
  Lowest order Feynman diagram
  describing \sel\ production in \ee\ collisions.
}
\label{fee}
\end{figure}

Selectron pair-production 
takes place in \ee\ collisions
via the exchange of \no s,
as depicted in the Feynman diagrams of Fig.~\ref{fee}.
The subsequent cascade decays of the \sel s 
into electrons and invisible particles
leads to the following observable signal:
\beq
\ee \quad\to\quad \tilde e^-\tilde e^- \quad\to\quad \ee+\mpT
\label{ee}~.
\eeq
This reaction and its \xs s have been discussed in details in Refs~\cite{ee}.

\begin{figure}[htb]
\begin{center}
\unitlength1mm
\SetScale{2.837}
\begin{picture}(50,50)(0,0)
\epsfig{file=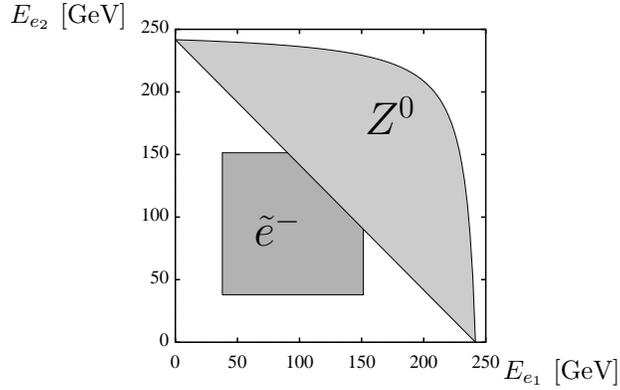,height=50mm}
\Text(-50,50)[tr]{\normalsize  $E_{e_2}$ [GeV]}
\Text(0,0)[bl]{\normalsize  $E_{e_1}$ [GeV]}
\Text(-30,20)[c]{\LARGE {$\tilde e^-$}}
\Text(-15,35)[c]{\LARGE {$Z^0$}}
\end{picture}
\end{center}
\caption{\footnotesize 
  Dalitz plot of the allowed energy ranges
  of the final state electrons in the processes
  $e^-e^-\to e^-e^-Z^0$
  and
  $e^-e^-\to\tilde e^-\tilde e^-\to e^-e^-\tilde\chi^0_1\tilde\chi^0_1$
  For the latter reaction
  we have assumed $m_{\tilde e}=$ 200 GeV
  and $m_{\tilde \chi^0_1}=$ 100 GeV.
}
\vskip-3mm
\label{dalitz}
\end{figure}

The most important \sm\ backgrounds 
originate from $W^-$ and $Z^0$ Brems\-strah\-lung
{\arraycolsep0cm
\renewcommand{\arraystretch}{0}
\beq
e^-e^- 
\quad\to\quad
\begin{array}[t]{ll}
  e^-\nu_e&W^- \qquad \\
  &\hra e^-\bar\nu_e
\end{array}
\qquad\qquad
e^-e^- 
\quad\to\quad
\begin{array}[t]{ll}
  e^-e^-&Z^0 \qquad \\
  &\strut\hra\nu\bar\nu
\end{array}
\label{zee}
\eeq
}
These backgrounds are higher order processes,
so that,
taking into account the heavier mass of the pair-produced \sel s,
one finds a signal to background ratio of the order of unity
over the whole \susy\ parameter space.
It is even possible to eliminate the \sm\ backgrounds
almost entirely.
This involves
{\em(i)}
	the use of right-polarized electron beams,
	in order to reduce  
	the $W^-$ Brems\-strah\-lung background
	to a negligible level;
{\em(ii)}
	rejecting all \ee\ events
	with a total deposited energy exceeding about half the \cm\ energy,
	in order to also filter out the $Z^0$ Bremsstrahlung events,
	as displayed in Fig.~\ref{dalitz}.

\begin{figure}[htb]
\vskip-10mm
\unitlength1mm\SetScale{2.837}
\begin{picture}(40,60)(0,0)
\Text(0,40)[tl]{\normalsize \fbox{$e_R^-e_R^- \to e^-e^-+\mpT$}}
\Text(0,30)[tl]{\normalsize $\sqrt{s_{ee}} = 500$ GeV}
\Text(0,25)[tl]{\normalsize ${\cal L}_{ee} = 10$ fb$^{-1}$}
\Text(0,20)[tl]{\normalsize $\tan\beta = 10$}
\Text(0,15)[tl]{\normalsize $m_{\tilde e} = 200$ GeV}
\end{picture}
\hskip15mm
\begin{picture}(50,50)(0,0)
\epsfig{file=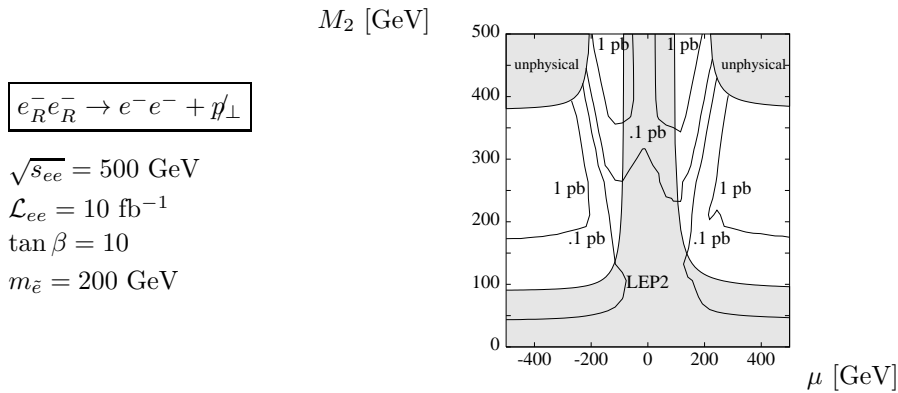,height=50mm}
\Text(-50,50)[tr]{\normalsize  $M_2$ [GeV]}
\Text(0,0)[bl]{\normalsize  $\mu$ [GeV]}
\end{picture}
\caption{\footnotesize
  Contours in the \susy\ parameter space
  of constant \xs s for the selectron signal.
}
\label{paramee}
\end{figure}

The next order irreducible background then
originates from double $W^-$ Bremsstrahlung 
{\arraycolsep0cm
\renewcommand{\arraystretch}{0}
\begin{equation}
        e^-e^- \quad\to\quad
        \begin{array}[t]{ll}
                W^-\nu_e&W^-\nu_e \qquad ,\\
                        &\hra e^-\bar\nu_e\\
                \multicolumn{2}{l}{\hra e^-\bar\nu_e}
        \end{array}
\label{wwee}
\end{equation}
}
and amounts to about .1 fb at 500 GeV~\cite{ckr}.
We have plotted in Fig.~\ref{paramee} 
the contours in the $(\mu,M_2)$ plane
along which the observable \xs\ for the signal (\ref{ee})
is 1 and 0.1 pb.

The nearly total absence of backgrounds
is a unique opportunity 
for performing studies
which would be arduous or impossible
in any other environment,
like \pe\ annihilations.
In particular,
the mass of the lightest \no\ can be very precisely determined
from the endpoints $E_{\rm min,max}$ 
of the electron energy distribution:
\bea
m_{\tilde\chi^0_1}^2
&=&
\sqrt{s}
{E_{\rm max}E_{\rm min} \over E_{\rm max}+E_{\rm min}}
\left( {\sqrt{s} \over E_{\rm max}+E_{\rm min}} -2 \right)
\label{lspee}\ .
\eea
This is a totally model-independent, 
kinematical measurement of the mass of the \lsp,
which no other experiment can perform 
within such a clean environment.

Moreover,
softer electrons emerging at the end
of a longer cascade
can also be observed.
This makes the \ee\ \lc\ mode
an ideal and unique tool
for observing and studying \susic\ cascades.
Neither hadronic nor \pe, \ep\ or \pp\ collisions
can perform well in this field,
because they all require cuts on the low \trm\ or energy signal electrons,
in order to enhance the signal to background ratio.

\section{\boldmath\ep\ Scattering \protect\cite{ep}}

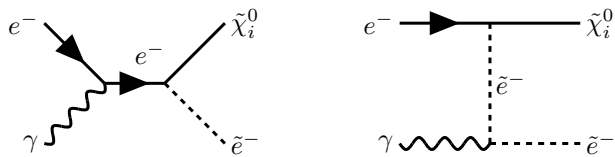
\begin{figure}[htb]
\unitlength.8mm\SetScale{2.269}
\begin{center}
\begin{picture}(40,20)(0,0)
\ArrowLine(0,20)(10,10)
\Text(-1,20)[r]{\normalsize $e^-$}
\Photon(0,0)(10,10){-1}{3.5}
\Text(-1,0)[r]{\normalsize $\gamma$}
\ArrowLine(10,10)(20,10)
\Text(15,15)[l]{\normalsize $e^-$}
\Line(20,10)(30,20)
\Text(31,20)[l]{\normalsize $\tilde\chi^0_i$}
\DashLine(20,10)(30,0){1}
\Text(31,0)[l]{\normalsize $\tilde e^-$}
\end{picture}
\qquad\qquad
\begin{picture}(40,20)(0,0)
\ArrowLine(0,20)(15,20)
\Text(-1,20)[r]{\normalsize $e^-$}
\Photon(0,0)(15,0){-1}{3.5}
\Text(-1,0)[r]{\normalsize $\gamma$}
\DashLine(15,20)(15,0){1}
\Text(16,10)[l]{\normalsize $\tilde e^-$}
\Line(15,20)(30,20)
\Text(31,20)[l]{\normalsize $\tilde\chi^0_i$}
\DashLine(15,0)(30,0){1}
\Text(31,0)[l]{\normalsize $\tilde e^-$}
\end{picture}
\end{center}
\caption{
  Lowest order Feynman diagram
  describing \sel\ production in \ep\ collisions.
}
\label{fep}
\end{figure}

Selectrons can be produced singly in \ep\ collisions,
as depicted in the Feynman diagrams of Fig.~\ref{fep}.
The subsequent decay (\ref{seldec}) of the \sel\ 
leads to the following observable signal:
\beq
\ep \quad\to\quad \tilde e^-\tilde\chi^0_1 \quad\to\quad e^-+\mpT
\label{ep}~.
\eeq
This reaction and its \xs s have been discussed in details in Refs~\cite{ep}.

\begin{figure}[htb]
\hskip25mm
\input{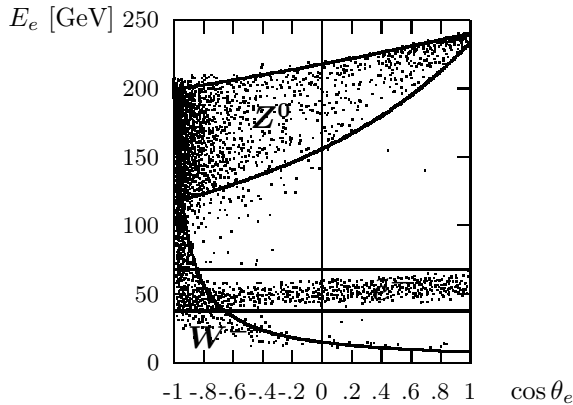}
\vskip-5mm
\caption{
	Electron's angle--energy distributions 
	in \ep\ scattering
	for a 250 GeV \sel\ decay
	and the backgrounds (\protect\ref{zep}).
	The collider's $e^\pm e^-$ energy and luminosity 
	are 500 GeV and 10 fb$^{-1}$.
        The curves show the exact $Z^0$
        and approximate $W^-$
        allowed areas.
}
\vskip-5mm
\label{cutsep}
\end{figure}

The most important \sm\ backgrounds 
originate from $W^-$ and $Z^0$ Brems\-strah\-lung
{\arraycolsep0cm
\renewcommand{\arraystretch}{0}
\beq
\ep
\quad\to\quad
\begin{array}[t]{ll}
  \nu_e&W^- \qquad \\
  &\hra e^-\bar\nu_e
\end{array}
\qquad\qquad
\ep
\quad\to\quad
\begin{array}[t]{ll}
  e^-&Z^0 \qquad \\
  &\strut\hra\nu\bar\nu
\end{array}
\label{zep}
\eeq
}
In contrast to the \ee\ reaction,
these backgrounds are very large
and more subtle cuts have to be imposed 
to obtain a good signal-to-background ratio.
The background reduction involves
{\em(i)}
	the use of right-polarized electron beams,
	in order to reduce
	the $W^-$ production background
	to an acceptable level;
{\em(ii)}
	imposing the kinematical cuts displayed in Fig.~\ref{cutsep},
	which entirely remove the $Z^0$ events
	and substantially suppress the remaining $W^-$ background;
{\em(iii)}
	enhancing the \susic\ signal
	by polarizing the laser and Compton-converted electron beams
	such as to have dominantly right-handed photons in the initial state;
{\em(iv)}
	dividing the $(\cos\theta_e,E_e)$ phase space 
	into $3\times3$ equal size bins,
	and computing a least squares estimator.

\begin{figure}[htb]
\unitlength1mm
\SetScale{2.837}
\hskip-0mm
\begin{picture}(40,0)(0,0)
\Text(0,45)[bl]{\normalsize \fbox{$e_R^-\gamma_R \to e^-+\mpT$}}
\Text(0,35)[bl]{\normalsize $\sqrt{s_{ee}} = 500$ GeV}
\Text(0,30)[bl]{\normalsize ${\cal L}_{ee} = 10$ fb$^{-1}$}
\Text(0,25)[bl]{\normalsize $\tan\beta = 10$}
\Text(0,20)[bl]{\normalsize $m_{\tilde e} =$}
\Text(9,20)[bl]{\normalsize $250$ GeV}
\Text(9,15)[bl]{\normalsize $300$ GeV}
\Text(9,10)[bl]{\normalsize $350$ GeV}
\Text(9,05)[bl]{\normalsize $400$ GeV}
\end{picture}
\input{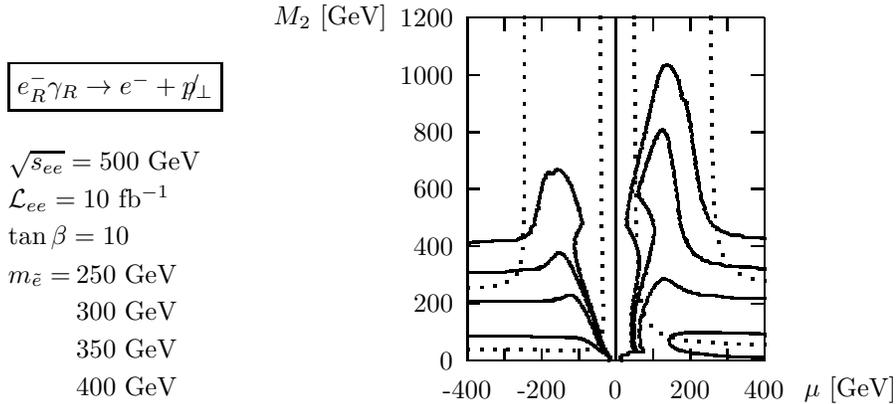}
\vskip-5mm
\caption{
	Contours of $\chi^2=6$ 
        for \sel\ masses, 
        increasing from the upper to lower curves.
        The dotted curves delimit the region already excluded by LEP I
        and the region below which charginos can be pair-produced 
        at the same machine run in the \pe\ mode.
}
\vskip-0mm
\label{paramep}
\end{figure}

In spite of the more complicated data analysis,
the advantage of \ep\ scattering
is the possibility of producing single \sel s.
In the event the collider energy is insufficient
to pair-produce them in \pe, \ee\ or \pp\ collisions,
heavy \sel s can still be observed in the \ep\ operating mode.
We have plotted in Fig.~\ref{paramep} 
the contours in the $(\mu,M_2)$ plane
of supersymmetry parameters
along which our $\chi^2$ estimator
provides a 95\%\ confidence 
for four \sel\ signal,
which are inaccessible via pair production.

As in \ee\ scattering,
the mass of the lightest \no\ can be determined kinematically
from the endpoints $E_{\rm min,max}$ 
of the electron energy distribution:
\bea
m_{\tilde\chi^0_1}^2
&=&
m_{\tilde e}^2 - 2m_{\tilde e} \sqrt{E_{\rm max}E_{\rm min}}
\label{lspep}\ .
\eea

\section{\boldmath\pp\ Scattering \protect\cite{pp}}

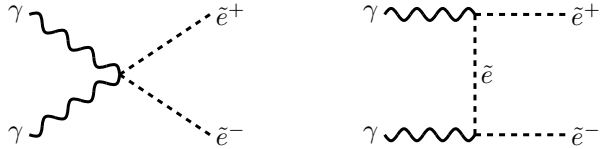
\begin{figure}[htb]
\unitlength.8mm\SetScale{2.269}
\begin{center}
\begin{picture}(40,20)(0,0)
\Photon(0,20)(15,10){1}{3.5}
\Text(-1,20)[r]{\normalsize $\gamma$}
\Photon(0,0)(15,10){-1}{3.5}
\Text(-1,0)[r]{\normalsize $\gamma$}
\DashLine(15,10)(30,20){1}
\Text(31,20)[l]{\normalsize $\tilde e^+$}
\DashLine(15,10)(30,0){1}
\Text(31,0)[l]{\normalsize $\tilde e^-$}
\end{picture}
\qquad\qquad
\begin{picture}(40,20)(0,0)
\Photon(0,0)(15,0){-1}{3.5}
\Text(-1,0)[r]{\normalsize $\gamma$}
\Photon(0,20)(15,20){1}{3.5}
\Text(-1,20)[r]{\normalsize $\gamma$}
\DashLine(15,20)(15,0){1}
\Text(16,10)[l]{\normalsize $\tilde e$}
\DashLine(15,20)(30,20){1}
\Text(31,20)[l]{\normalsize $\tilde e^+$}
\DashLine(15,0)(30,0){1}
\Text(31,0)[l]{\normalsize $\tilde e^-$}
\end{picture}
\end{center}
\caption{
  Lowest order Feynman diagram
  describing \sel\ production in \pp\ collisions.
}
\label{fpp}
\end{figure}

Selectron pair-production 
takes place in \pp\ collisions
as depicted in the Feynman diagrams of Fig.~\ref{fpp}.
The polarized \xs s have the following
threshold and asymptotic behaviours \cite{pp}:
\bea
\label{threshpp}
m_{\tilde e}\approx{\sqrt{s}\over2}: &&
\sigma = {\pi\alpha^2\over s} 
\left( 1-P_1P_2 \right)
\sqrt{1-{4m_{\tilde e}^2\over s}}
\\
\label{asympp}
m_{\tilde e}\ll\sqrt{s}: &&
\sigma = {\pi\alpha^2\over s}
\left( 1+P_1P_2 \right)~.
\eea
The subsequent decay (\ref{seldec}) of the \sel s 
leads to the following observable signal:
\beq
\pp \quad\to\quad \tilde e^+\tilde e^- \quad\to\quad \pe+\mpT
\label{pp}~.
\eeq

\begin{figure}[htb]
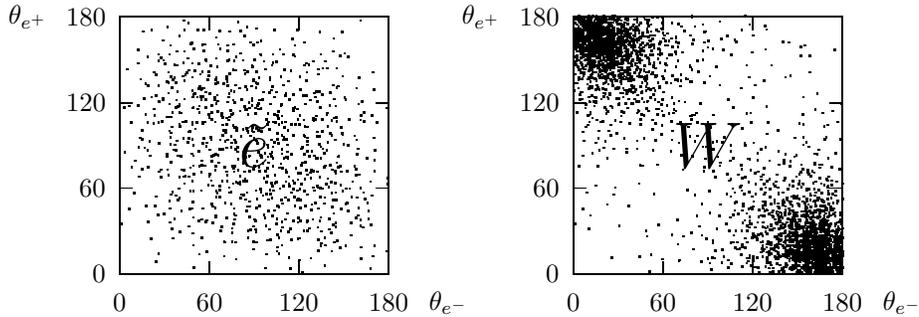

\hskip-9mm\input{scats.tex}
\hskip6mm\input{scatw.tex}
\vskip-10mm
\caption{
  Angular distributions of the electrons emerging from the decays of 
  300 GeV \sel s and 
  the background (\protect\ref{wwpp}).
  The collider's $e^\pm e^-$ energy and luminosity are 1 TeV and 80 fb$^{-1}$.
}
\label{scat}
\end{figure}

There are many backgrounds to this signal.
Most of them,
though,
can easily be eliminated by mild detector acceptance cuts.
However,
the only one which survives these cuts
{\arraycolsep0cm
\renewcommand{\arraystretch}{0}
\begin{equation}
        \pp \quad\to\quad
        \begin{array}[t]{ll}
                W^+&W^- \\
                        &\hra e^-\bar\nu_e\\
                \multicolumn{2}{l}{\hra e^+\nu_e}
        \end{array}
\label{wwpp}
\end{equation}
}
is important and has to be suppressed.
The background reduction involves
{\em(i)}
using photon beams 
with opposite helicity in the threshold region
or same helicity at asymptotic energies
({\em cf.} Eqs~(\ref{threshpp},\ref{asympp}));
{\em(ii)}
cutting out all events 
which cannot originate from \sel\ production and decay,
{\em i.e.}, 
for which the energy of the electrons
is not confined within
\begin{equation}
        E_e  \in {E\over4} \left[1-{m^2_{\tilde\chi^0_1}\over m^2_{\tilde e}}\right]
                \left[1\pm\sqrt{1-{4m^2_{\tilde e}\over E^2}}\right]
\qquad
E\simeq.83\sqrt{s_{ee}}
\label{elep}
\end{equation}
(the \sel\ mass having undoubtedly been previously determined 
in \ee\ or \pe\ experiments);
{\em(iii)}
dividing the angular phase space,
whose characteristic signal and background demography 
are displayed in Figs~\ref{scat},
into $3\times3$ equal size bins,
and computing a least squares estimator.

\begin{figure}[htb]
\hskip0mm
\unitlength1mm
\SetScale{2.837}
\begin{picture}(40,50)(0,0)
\Text(0,45)[bl]{\normalsize \fbox{$\gamma_L\gamma_R \to e^+e^-+\mpT$}}
\Text(0,35)[bl]{\normalsize $\sqrt{s_{ee}} = 1$ TeV}
\Text(0,30)[bl]{\normalsize ${\cal L}_{ee} = 80$ fb$^{-1}$}
\Text(0,25)[bl]{\normalsize $\tan\beta = 4$}
\Text(0,20)[bl]{\normalsize $m_{\tilde e} =$}
\Text(9,20)[bl]{\normalsize $300$ GeV}
\Text(9,15)[bl]{\normalsize $350$ GeV}
\end{picture}
\input{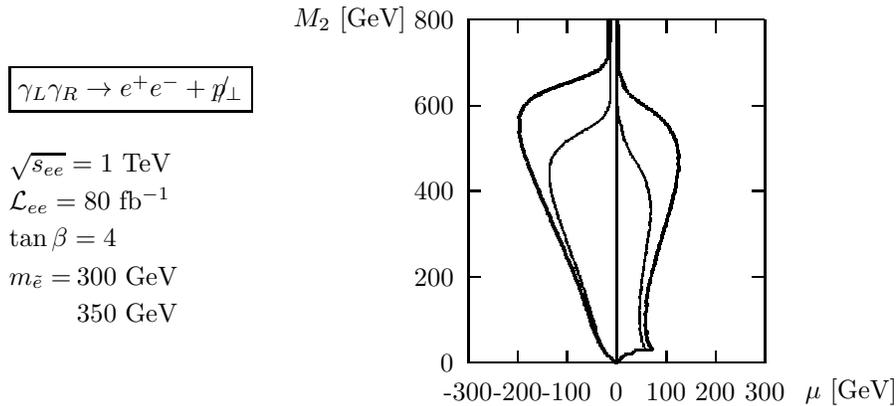}
\vskip-5mm
\caption{
  Contours of $\chi^2=6$.
  The \sel\ masses are $m_{\tilde e}=300,350$ GeV,
  for the inner and outer curves respectively.
}
\vskip-5mm
\label{cont1}
\end{figure}

Although for discovering \sel s
\pp\ collisions are no challenge to the other modes 
they allow for a direct measurement of the \br\ for \sel\ decays.
Indeed,
once the \sel\ mass is determined,
the total production \xs s are uniquely predicted.
Demanding $\chi^2=6$,
one obtains in Fig.~\ref{cont1} 
contours in the $(\mu,M_2)$ plane,
which are barely distinguishable
from the curves delimiting the regions of parameter space
where the branching ratio of the decay (\ref{seldec})
is .02 or .035
(for \sel s of mass 300 and 350 GeV respectively).
Clearly,
this direct measurement of the \sel\ branching ratios
will provide invaluable information 
on the values of the \susy\ parameters.
Obviously,
the same analysis applies also for smuons,
andto a less clean extent to staus.

\section{Conclusions}

The \ee, \ep\ and \pp\ modes of a \lc\
provide powerfull experiments,
which are complementary to each other and to \pe\ collisions.
We have summarized here 
the results of the \susy\ analysis performed in these operating modes.

The \ee\ mode
benefits from its low \sm\ background activity
and is capable of discovering 
any kinematically accessible \sel\
by a simple counting experiment.
Moreover,
an extremely pure sample of right-\sel s can be obtained
with polarized beams,
which allows 
the precise measurement of the lightest \no\ mass and
the opportunity to observe and analyze cascade decays of the \sel.

If the collider energy is insufficient
to pair-produce \sel s,
they may still be produced singly with the \ep\ option.
Here again,
the mass of the lightest \no\ 
can be determined in a model-independent way,
but not as precisely as in \ee\ scattering,
because of the incomplete background reduction.

The \pp\ mode is not suited for discovering \sel s,
because of its inherent lower \cm\ energy.
Nevertheless,
it provides a direct, model-independent measurement 
of the \br\
of \sel's main decay mode.
This information can be advantageously used 
to further constrain the \susy\ parameter space.

\end{document}